\documentclass[doublecol]{epl2}
\usepackage[T1]{fontenc}
\usepackage{lmodern}
\usepackage{amsmath,amssymb}
\usepackage{xurl}
\usepackage{booktabs}
\usepackage{hyperref}

\newcommand{\dd}{\mathrm{d}}
\newcommand{\A}{\rm an}
\newcommand{\Q}{\rm qu}
\newcommand{\p}{{\rm Pe}}
\newcommand{\average}[1]{\left\langle #1 \right\rangle}

\makeatletter
\def\dsh@theta#1{%
  \ifx#1\displaystyle      \fontdimen8\textfont\thr@@
  \else\ifx#1\textstyle    \fontdimen8\textfont\thr@@
  \else\ifx#1\scriptstyle  \fontdimen8\scriptfont\thr@@
  \else                    \fontdimen8\scriptscriptfont\thr@@ \fi\fi\fi}
\newcommand{\dashoverline}[1]{\mathpalette\dsh@over{#1}}
\newcommand{\dsh@over}[2]{%
  \sbox\z@{$\m@th#1#2$}%
  \dimen@=\dsh@theta{#1}%
  \vbox{\offinterlineskip
    \kern\dimen@
    \hbox to\wd\z@{\xleaders\hbox{\kern2.5\dimen@\vrule height1.5\dimen@ width7\dimen@\kern2.5\dimen@}\hfil}%
    \kern2\dimen@
    \box\z@}}
\makeatother

\title{Current fluctuations in a gas of active Ornstein--Uhlenbeck particles}
\shorttitle{Current fluctuations in active Ornstein--Uhlenbeck particles}

\author{Sandeep Jangid \inst{1} \and Aman Kumbhakar \inst{1} \and Juliane U. Klamser \inst{2} \and Tridib Sadhu \inst{1}}
\shortauthor{S. Jangid \etal}

\institute{
  \inst{1} Department of Theoretical Physics, Tata Institute of Fundamental Research, Homi Bhabha Road, Mumbai 400005, India.\\
  \inst{2} Laboratoire Charles Coulomb (L2C), Universit\'e de Montpellier and CNRS (UMR 5221), 34095 Montpellier, France.
}

\abstract{We investigate the statistics of the time-integrated current in an infinite one-dimensional gas of independent active Ornstein--Uhlenbeck particles, as a model system for studying an active generalisation of the corresponding passive (diffusive) phenomenology. Unlike the latter, where current fluctuations exhibit universal sub-diffusive scaling, active systems display diffusive, super-diffusive, and sub-diffusive regimes over different time scales. We fully characterise the distribution of current in terms of large-deviation asymptotics, showing that all of these scaling regimes are described by a single scaled cumulant generating function. Moreover, the statistics retain a dependence on the initial condition even at large times, revealing a persistent memory of the initial state. We further obtain the joint large-deviation statistics of currents measured at two distinct times, characterising temporal correlations. Our analytical predictions are verified using rare-event importance sampling, which resolves probabilities as small as $10^{-1000}$.
}

\begin{document}
\maketitle

\section{Introduction}

Current fluctuations provide a direct probe of transport properties of many-body systems and have been extensively studied in both classical and quantum settings~\cite{2015_Lazarescu_The,2002_Prahofer_Current,2004_bodineau_Current,1995_Lee_Universal,2023_McCulloch_Full,2024_Wienand_Emergence}. Decades of work on current fluctuations in ballistic and diffusive dynamics have uncovered emergent fluctuation symmetries~\cite{1999_lebowitz_gallavotti,2004_Roche,2011_hurtado_symmetries}, universal correlations~\cite{2015_popkov_fibonacci,2023_benjamin_emergence,2016_sadhu_correlations}, and linear-response relations~\cite{1966_kubo_fluctuation,2016_sadhu_correlations,2009_marco_fluctuations}. The efforts to understand them have driven the development of fluctuating hydrodynamics~\cite{Bertini2014,2014_herbert_nonlinear,2026_Saha_Bottom} and of integrability-based methods~\cite{2007_Derrida_Non,2022_Mallick_Exact}. By contrast, the corresponding phenomenology for overdamped active dynamics, which are ballistic at short scales and diffusive beyond, remains surprisingly little understood~\cite{2020_Banerjee_Current,2023_Bello_Current,2023_Jose_Generalized,2024_Jose_Effect}.

Active matter constitutes an important class of nonequilibrium systems, in which individual constituents continuously consume energy to generate systematic motion~\cite{ramaswamy2010mechanics,2022_bowick_symmetry,Fodor2018}. This microscopic breaking of detailed balance gives rise to a wide variety of collective and dynamical phenomena with no direct equilibrium counterpart~\cite{1995_vicsek_Novel,2015_cates_motility,2016_Bechinger}. Despite extensive progress in active-matter physics, quantitative studies of transport properties and fluctuations have received comparatively little attention~\cite{2020_Banerjee_Current,2023_Bello_Current,2023_Jose_Generalized,2024_Jose_Effect,2023_agranov_macroscopic}. A key open question is how the universal characteristics of fluctuations in passive many-body transport generalise for active dynamics. In this \emph{Letter}, we address this question by studying current fluctuations in a model active-matter system. 

Current fluctuations are traditionally characterised using the tools of full-counting statistics~\cite{1995_Lee_Universal} and large-deviation theory~\cite{Touchette2009,2025_Bernard_Lecture}. Beyond providing a complete statistical description at the macroscopic scale, these statistics serve to characterise the thermodynamics of nonequilibrium many-body systems. Yet even for passive many-body systems, determining current statistics remains a significant challenge, and only a handful of exact results are known~\cite{Lecomte2010,Akkermans2013,2015_Lazarescu_The,Bodineau2005,2009_Derrida_Current,2022_Mallick_Exact, 2024_kapil_semi,2026_Saha_Effect,2026_Jangid_An,2026_Kethepalli_Ballistic,2026_saha_universal}. 

For active dynamics, the only available equivalent results are for one-dimensional, non-interacting run-and-tumble particles (RTPs)~\cite{2020_Banerjee_Current,2023_Bello_Current,2023_Jose_Generalized,2023_jose_current,2024_chakraborty_current} and active lattice gases~\cite{2023_agranov_macroscopic,2024_Jose_Effect}. These pioneering works reveal clear departures from the passive phenomenology. In particular, the variance of the current exhibits distinct scaling regimes~\cite{2023_Jose_Generalized}: linear for times below the persistence time (as in passive ballistic transport~\cite{2026_Kethepalli_Ballistic,2026_saha_universal}) and $\propto\sqrt{t}$ at large times (as in passive diffusive systems~\cite{2009_Derrida_Current2}). Moreover, initial conditions drastically alter the time scaling~\cite{2023_Jose_Generalized}, which is in stark contrast to passive systems~\cite{2009_Derrida_Current,2026_saha_universal}. Beyond these few dynamical characteristics, a full statistical characterisation for generic initial conditions has proved challenging even in the non-interacting case~\cite{2020_Banerjee_Current} and the multi-time statistics probing temporal correlations remain largely unexplored.

In this \emph{Letter}, we study the current fluctuations, including their multi-time statistics, in an infinite one-dimensional gas of non-interacting active Ornstein--Uhlenbeck particles (AOUPs)~\cite{2021_Martin_statistical}. AOUPs, along with RTPs and active Brownian particles (ABPs), constitute the three minimal models of self-propulsion dynamics without alignment interactions. Unlike RTPs and ABPs, the dynamics of AOUPs is inherently Gaussian, which allows us to fully characterise current statistics and their sensitivity to initial conditions.

Our explicit results show that the distinct scaling regimes of current fluctuations are described by a single scaled cumulant generating function (SCGF), which determines cumulants of all orders. The SCGF, and the scaling with time, both depend on the choice of the initial state, demonstrating that the initial condition affects not only the numerical prefactors (as in passive systems~\cite{2026_saha_universal}) but also the temporal structure of the fluctuations and how they relate to the single-particle statistics. We further extend our analysis beyond single-time statistics to characterise temporal correlations of the integrated current. We verify our results using rare-event simulations~\cite{2023_Bello_Current}, which probe probabilities as low as $10^{-1000}$. 

\begin{figure}
    \centering
    \includegraphics[width=0.9\linewidth]{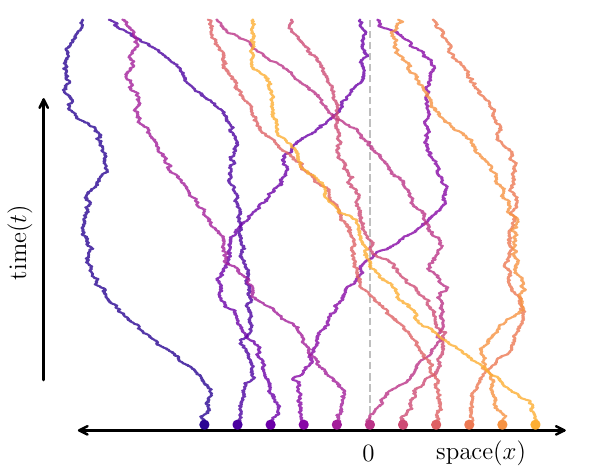}
    \caption{Schematic trajectories of a collection of non-interacting AOUPs on a one-dimensional line. In this example, all particles start with zero initial velocity ($v_0=0$). Particle current is measured across the origin, indicated by a dashed line.}
    \label{fig:Trajectories}
\end{figure}

\section{Model}

\begin{subequations}
We consider a gas of independent self-propelled particles on an infinite line (see Fig.~\ref{fig:Trajectories}). The position of each particle evolves as
\begin{equation}\label{eq: langevin equation x}
    \dot{x}_t = v_t + \sqrt{2 D}\, \eta_t
\end{equation}
where $v_t$ is the self-propulsion velocity, $D$ is the bare thermal diffusivity, and $\eta_t$ is a Gaussian white noise with zero mean and unit strength. The self-propulsion velocity follows the AOUP dynamics~\cite{2021_Martin_statistical} 
\begin{equation}\label{eq: langevin equation v}
    \tau \dot{v}_t = -v_t + \p \sqrt{2 D} \,\xi_t
\end{equation}
where $\tau$ is the persistence time, $\p$ is the P\'eclet number, and $\xi_t$ is an independent Gaussian white noise with the same statistics as $\eta_t$. The coefficient $\p$ in Eq.\eqref{eq: langevin equation v} is chosen so that it coincides with the standard P\'eclet number $\hat{v} \tfrac{\ell_d}{D}$, where $\hat{v} = \sqrt{\langle v_t^2\rangle}$ is the resulting stationary speed and $\ell_d = \sqrt{D\tau}$. The system is initially prepared in a domain-wall state where the particles are randomly distributed in position with average density $\rho_a$ on the left of the origin and $\rho_b$ on the right. The initial velocities of particles are drawn independently from a Maxwellian distribution
\begin{equation}\label{eq: Maxwell Boltzmann}
    p(v_0=u) =  \sqrt{\frac{\tau}{2 \pi \p^2 D}} \mathrm{e}^{- \frac{\tau u^2}{2 \p^2 D}}
\end{equation}
which is the stationary distribution of Eq.~\eqref{eq: langevin equation v}.
\end{subequations}

\section{Setup}

The central observable of interest is the time-integrated current $Q_t$, which represents the total particle flux across the origin over a time period $t$. The corresponding full-counting statistics is characterised in terms of the generating function $\langle e^{\lambda Q_t}\rangle$, where $\lambda$ is the fugacity parameter. To quantify the sensitivity to initial conditions we borrow the idea of annealed and quenched averages of the free energy in disordered systems~\cite{2011_Berthier_Theoretical,2009_Derrida_Current2,2015_sadhu_large}. In our context, the partition function is the generating function, and in this analogy we define the cumulant generating function (CGF) according to the ensemble of initial fluctuations:
\begin{subequations}\begin{align}
 \mu_{\A,\A}(\lambda)&=\ln\overline{\dashoverline{\langle e^{\lambda Q_t}\rangle}}&
 \mu_{\A,\Q}(\lambda)&=\dashoverline{\ln\overline{\langle e^{\lambda Q_t}\rangle}}\\
 \mu_{\Q,\A}(\lambda)&=\overline{\ln\dashoverline{\langle e^{\lambda Q_t}\rangle}}&
 \mu_{\Q,\Q}(\lambda)&=\dashoverline{\overline{\ln\langle e^{\lambda Q_t}\rangle}}
\end{align}\label{eq:ensemble_defs}\end{subequations}
where $\left<\cdot\right>$ denotes averaging over the time evolution, while $\overline{\cdots}$ and $\dashoverline{~\cdots ~}$ represent averaging over the initial positions and initial velocities of the particles, respectively. The subscripts `$\A$' and `$\Q$' refer to the annealed and quenched ensembles, respectively, with the first subscript referring to position and the second to velocity. We also consider an ensemble in which all initial velocities are precisely zero and denote the corresponding CGF by $\mu_{\Q,0}(\lambda)=\overline{\ln\langle e^{\lambda Q_t}\rangle}_{u=0}$. 

\section{Main results}\label{sec:results}

The simplest case is the annealed average of the initial positions. Irrespective of the choice of average (annealed or quenched) over initial velocities, the CGF takes the form
\begin{equation}
     \mu_{\A,\A(\Q)}(\lambda)
     = \frac{\tilde\sigma_t}{\sqrt{2\pi}}\left(\rho_a(e^\lambda-1)+\rho_b(e^{-\lambda}-1)\right)
     \label{eq: result aa}
\end{equation}
resembling the corresponding result for the passive case~\cite{2009_Derrida_Current2}. The time dependence is in the proportionality factor
\begin{equation}\label{eq: tilda_sigma}
        \tilde{\sigma}_t^2 = 2Dt\left(1+\p^2\right)-2 D \tau \p^2\left(1-e^{-\frac{t}{\tau}}\right)
\end{equation}
and Eq.~\eqref{eq: result aa} implies that all cumulants scale with $\tilde{\sigma}_t$.

\begin{figure*}
    \centering
    \includegraphics[width=1\linewidth]{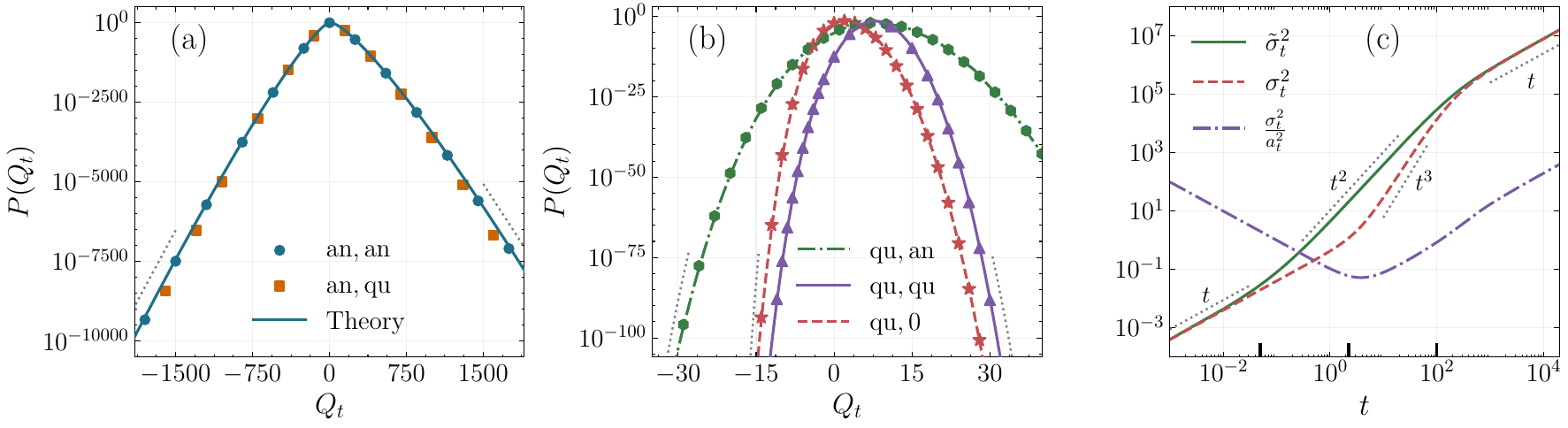}
    \caption{\emph{Probability distribution of the current:} (a) For annealed initial positions, the circles and squares denote the numerical simulation results obtained by importance sampling~\cite{Supp_Mat} for the annealed-velocity and quenched-velocity ensembles, respectively. The blue solid line represents the normalised distribution in Eq.~\eqref{eq:ldf scaling} with $\Phi(q)$ obtained by Legendre transform of Eq.~\eqref{eq: result aa}. The dotted lines indicate the $e^{-Q_t \ln Q_t}$-tails. (b) For quenched initial positions, the green circles, purple triangles, and red stars denote numerical simulation results for the annealed-velocity, quenched-velocity, and zero-velocity ensembles, respectively. The green dash-dotted, purple solid, and red dashed lines are the corresponding theoretical results obtained from Eq.~\eqref{eq:ldf scaling}. The dotted lines indicate the $e^{-Q_t^3}$-tails. Both (a) and (b) are for $\rho_a = \tfrac{3}{2}$ and $\rho_b = \tfrac{1}{2}$, and $t = 10$. (c) The time dependence of the scaling parameters $\tilde{\sigma}_t^2$ and ${\sigma}_t^2$ in Eqs.~(\ref{eq: tilda_sigma},\ref{eq: sigma}) shows three distinct dynamical regimes. The vertical thick black bars on the $x$-axis indicate $\tfrac{\tau}{\p^2}$, $\tfrac{\tau}{\p}$, and $\tau$, in increasing order. Also shown is $\tfrac{\sigma_t^2}{a_t^2}$, which enters $\mu_{\Q,\Q}(\lambda)$ in Eq.~\eqref{eq: 1t_mu_QQ}. All figures are for $D = 0.2$, $\tau = 100 $, and $ \p = 44.7$.}
    \label{fig:one_time_probability_distribution}
\end{figure*}

The choice of the initial velocity average matters when the positions are quenched. For annealed velocities,
    \begin{align}
         \mu_{\Q,\A} (\lambda) = \tilde\sigma_t \,\sqrt2 &\int_0^\infty \dd z\bigg\{\rho_a
         \ln \left(1+\frac{e^\lambda-1}{2}\mathrm{erfc}(z)\right)\notag\\
         &+\rho_b
         \ln\!\left(1+\frac{e^{-\lambda}-1}{2}\mathrm{erfc}(z)\right)\!\bigg\}\label{eq: result mu_qa}
    \end{align}
with $\mathrm{erfc}$ being the complementary error function, while for quenched velocities,
\begin{align}\label{eq: 1t_mu_QQ}
         \mu_{\Q,\Q}&(\lambda)
         =\frac{\sigma_t}{\sqrt2} \int_{-\infty}^{\infty}\dd z
         \bigg\{\rho_a\ln\!\left(e^\lambda+\frac{1-e^\lambda}{2}\mathrm{erfc}(z)\right)\notag\\ 
         +&\rho_b\ln\left(e^{-\lambda}+\frac{1-e^{-\lambda}}{2}\mathrm{erfc}(z)\right)\bigg\}   \,\mathrm{erfc} \left(\frac{ z \sigma_t}{a_t}\right)
\end{align}
where the time dependence is through parameters
\begin{equation}\label{eq: sigma}
        \sigma_t^2= \tilde{\sigma}_t^2-a_t^2\quad\textrm{and } a_t=\p \sqrt{D\tau}\left(1-e^{-\tfrac{t}{\tau}}\right)
\end{equation}

The most noticeable difference is in the time-dependent prefactor: $\tilde{\sigma}_t$ in Eq.~\eqref{eq: result mu_qa}, as opposed to $\sigma_t$ in Eq.~\eqref{eq: 1t_mu_QQ}. These two prefactors reflect distinct scaling behaviour in time. In Eq.~\eqref{eq: result mu_qa}, all cumulants of the current scale as $\tilde{\sigma}_t$, whereas in Eq.~\eqref{eq: 1t_mu_QQ}, the even cumulants scale with $\sigma_t$, while the odd cumulants scale as $\sigma_t c_n (\tfrac{\sigma_t}{a_t})$ with the scaling function $c_n(x)$ deduced from Eq.~\eqref{eq: 1t_mu_QQ}. 

For large $\tilde{\sigma}_t$, these scalings in Eqs.~(\ref{eq: result aa},\ref{eq: result mu_qa}) give a large-deviation asymptotic form for the distribution of the current
\begin{equation}\label{eq:ldf scaling}
    P(Q_t) \sim \mathrm{e}^{-r_t \Phi\big(\frac{Q_t}{r_t}\big)}
\end{equation}  
with $r_t\equiv \tilde{\sigma}_t$, where the large-deviation function (LDF) $\Phi(q)$ relates to the corresponding CGFs via a Legendre transformation~\cite{2007_Derrida_Non,2009_Derrida_Current2}. Similarly, for equilibrium initial states ($\rho_a=\rho_b$), Eq.~\eqref{eq: 1t_mu_QQ} yields an analogous asymptotic form with $r_t\equiv \sigma_t$ taken large. For the non-uniform case, a similar asymptotic for holds for large $\sigma_t$, with $\tfrac{\sigma_t}{a_t}$ held fixed.  
We present an extensive numerical confirmation of Eq.~\eqref{eq:ldf scaling} in Fig.~\ref{fig:one_time_probability_distribution} using rare-event simulations~\cite{Supp_Mat}.

The distribution in Eq.~\eqref{eq:ldf scaling} gives $\sqrt{r_t}$ as the scale of typical current fluctuations. By contrast, in the far tails of positive rare fluctuations, $P(Q_t)\sim e^{-Q_t\log Q_t}$ for the annealed-position ensemble, irrespective of the velocity ensemble, while $P(Q_t)\sim e^{-Q_t^3}$ (ignoring numerical factors) for the remaining ensembles. Both are obtained from the asymptotics of the corresponding CGFs and are illustrated in Fig.~\ref{fig:one_time_probability_distribution}. 

More generally, all three dynamical regimes of active particles are visible in Eq.~\eqref{eq:ldf scaling} through the explicit time dependence of the scaling parameters, shown in Fig.~\ref{fig:one_time_probability_distribution}(c). The diffusive scaling $\tilde{\sigma}_t^2\sim t$ and $\sigma_t^2\sim t$ at both short and long times is characteristic of active particles with thermal diffusion, while persistent motion dominates at intermediate times. The ballistic scaling $\tilde{\sigma}_t^2\sim t^2$ holds between the time scales $\tfrac{\tau}{\p^2}$ and $\tau$, while $\sigma_t^2\sim t^3$ holds between $\tfrac{\tau}{\p}$ and $\tau$. We will show that $\sigma_t^2$ and $\tilde{\sigma}_t^2$ are in fact the mean-square displacement of a single AOUP, for zero initial velocity and averaged over the initial velocity distribution~\eqref{eq: Maxwell Boltzmann}, respectively. 

Perhaps the most interesting aspect that emerges from these exact results is that even the long-time statistics is sensitive to the details of initial configurations. This is because annealed averaging can be dominated by rare, atypical configurations, while quenched averaging is governed by the typical configuration alone. Our results show when such rare initial configurations become relevant.
Surprisingly, and against the conventional expectation from passive systems~\cite{Leibovich2013,KMS_JSP}, the quenched average is not always equivalent to fixing the initial state to a specific typical configuration. This is demonstrated by our result
\begin{equation}
     \mu_{\Q,0}(\lambda)=\frac{\sigma_t}{\tilde\sigma_t}\mu_{\Q,\A}(\lambda)
    \label{eq: result mu_q, u=0}
\end{equation}
which differs from the CGF in Eq.~\eqref{eq: 1t_mu_QQ}. Interestingly, the two CGFs coincide in the equilibrium state ($\rho_a=\rho_b$), where $a_t$ dependence in Eq.~\eqref{eq: 1t_mu_QQ} vanishes by the odd parity of the integrand.

Another important aspect is that the current fluctuations in the annealed-position ensemble obey a symmetry following directly from Eq.~\eqref{eq: result aa}:
\begin{equation}
    \mu_{\A, \A(\Q)}(\lambda) = \mu_{\A, \A(\Q)}\left(-\lambda-\ln \frac{\rho_a}{\rho_b}\right)
\end{equation}
analogous to the Gallavotti--Cohen--Evans--Morriss--Searles symmetry known for passive dynamics~\cite{2004_Roche,1999_lebowitz_gallavotti}.

We note that the corresponding statistics for RTPs remain considerably less complete~\cite{2020_Banerjee_Current,2023_Jose_Generalized,2024_Jose_Effect}. By contrast, for AOUPs, results are straightforward to obtain even for multi-time statistics of the current, revealing a long-range nature of the dynamical correlations that is characteristically different from the passive counterpart. As an explicit example, we consider the two-time CGF,
\begin{equation}
    \mu(\lambda_1, \lambda_2) = \ln \left\langle \mathrm{e}^{\lambda_1 Q_{t_1} + \lambda_2 Q_{t_2}}\right\rangle
\end{equation}
where we restrict ourselves to $t_2 \ge t_1$. Different initial ensembles are defined analogously to Eq.~\eqref{eq:ensemble_defs}.

\begin{figure}
    \centering
    \includegraphics[width=1.0\linewidth]{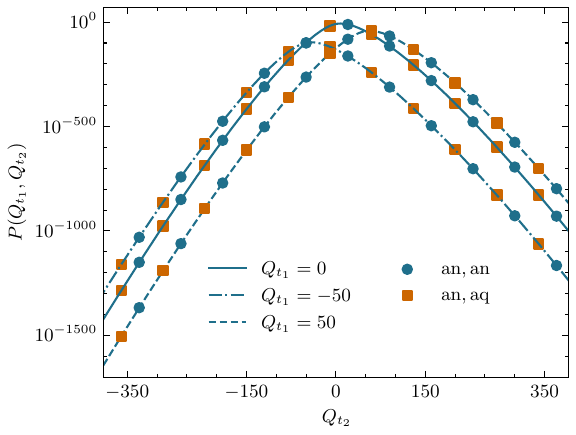}
    \caption{\emph{Joint distribution of the current at two times:} The lines show the joint probability distribution $P(Q_{t_1}, Q_{t_2})$, plotted as a function of $Q_{t_2}$ for fixed $Q_{t_1}$, obtained using the Legendre transform of Eq.~\eqref{eq: 2t_mu_AA} and writing the distribution in a form analogous to Eq.~\eqref{eq:ldf scaling}. The results shown are for times $t_1 = 10$ and $t_2 = 20$, with densities $\rho_a = \frac{3}{2}$ and $\rho_b = \frac{1}{2}$. The symbols denote the corresponding results from rare-event simulations. The remaining parameters are the same as in Fig.~\ref{fig:one_time_probability_distribution}.}
    \label{fig:two_time_AA_AQ}
\end{figure}

For the annealed-position, annealed-velocity ensemble, the two-time CGF has a simple expression:
\begin{subequations}\label{eq: 2t_mu_AA}
\begin{equation}
        \mu_{\A,\A}(\lambda_1,\lambda_2) = \mu_{\A,\A} (\lambda_1) + \mu_{\A,\A} (\lambda_2) + \nu (\lambda_1,\lambda_2)
\end{equation}
where the coupling term
\begin{align}\label{eq: two-time psi}
    &\nu(\lambda_1,\lambda_2) = \left(\tilde{\sigma}_{t_1} + \tilde{\sigma}_{t_2} - \tilde{\sigma}_{t_2-t_1}\right) \frac{1}{\sqrt{8\pi}} \notag \\
    & \Big(\rho_a(e^{\lambda_1}-1)(e^{\lambda_2}-1) + \rho_b(e^{-\lambda_1}-1)(e^{-\lambda_2}-1)\Big)
\end{align}
\end{subequations}
Setting either $\lambda_1$ or $\lambda_2$ to zero recovers Eq.~\eqref{eq: result aa}. As in Eq.~\eqref{eq: result aa}, we find $\mu_{\A,\Q} (\lambda_1,\lambda_2) = \mu_{\A,\A} (\lambda_1,\lambda_2)$. Both CGFs are verified against rare-event simulations in Fig.~\ref{fig:two_time_AA_AQ}. 

The corresponding CGFs for the remaining ensembles are more cumbersome, and we present them in the Supplemental Material~\cite{Supp_Mat}. Instead, here we present explicit results for the two-time correlations obtained from a series expansion of the corresponding CGFs. These correlations highlight distinct dynamical characteristics of the different ensembles.

For annealed positions, we obtain from Eq.~\eqref{eq: 2t_mu_AA} 
\begin{subequations}\label{eq:correlations}
\begin{equation}\label{eq: correlation AA}
    \langle Q_{t_1} Q_{t_2} \rangle_{\A, \A(\Q)} = \Gamma_{ab} \left(\tilde{\sigma}_{t_1} + \tilde{\sigma}_{t_2} - \tilde{\sigma}_{t_2-t_1}\right)
\end{equation}
with the prefactor $\Gamma_{ab} = \frac{\rho_a + \rho_b}{\sqrt{8\pi}}$. In the three dynamical regimes of $\tilde\sigma_t^2$ [Fig.~\ref{fig:one_time_probability_distribution}(c)], the time dependence of Eq.~\eqref{eq: correlation AA} takes the form of the covariance of fractional Brownian motion~\cite{2018_Tridib_Generalized,2021_Sadhu_Functionals}, with Hurst exponents $\tfrac{1}{4}$, $\tfrac{1}{2}$, and $\tfrac{1}{4}$, respectively.

For the quenched-position, annealed-velocity ensemble, the correlation reads~\cite{Supp_Mat}
\begin{equation}\label{eq: correlation QA}
        \langle Q_{t_1} Q_{t_2} \rangle_{\Q, \A} = \Gamma_{ab} \left( \sqrt{\tilde{\sigma}_{t_1}^2 + \tilde{\sigma}_{t_2}^2} - \tilde{\sigma}_{t_2-t_1} \right)
\end{equation}
which, in the same three dynamical regimes of $\tilde\sigma_t^2$ [Fig.~\ref{fig:one_time_probability_distribution}(c)],
resembles the correlations of the corresponding passive quenched dynamics~\cite{2026_saha_universal}.

The remaining velocity ensembles exhibit current correlations with no analogue in passive systems. For the $(\Q,\Q)$ ensemble,
\begin{equation}\label{eq: correlation QQ}
    \langle Q_{t_1} Q_{t_2} \rangle_{\Q,\Q} = \Gamma_{ab} \Big( \sqrt{\sigma_{t_1}^2 + \sigma_{t_2}^2 + \gamma_{12}^2} - \tilde{\sigma}_{t_2-t_1}\Big)
\end{equation}
with $\gamma_{12}^2 = (a_{t_1} - a_{t_2})^2$, while, for the $(\Q,0)$ ensemble,
\begin{equation}\label{eq: correlation Q0}
    \langle Q_{t_1} Q_{t_2} \rangle_{\Q,0} = \Gamma_{ab} \Big( \sqrt{\sigma_{t_1}^2 + \sigma_{t_2}^2} - \sqrt{\tilde{\sigma}^2_{t_2-t_1} - \gamma_{12}^2}\Big)
\end{equation}
See~\cite{Supp_Mat} for the corresponding CGFs.
Note that Eqs.~\eqref{eq: correlation QQ} and~\eqref{eq: correlation Q0} involve both $\sigma_t$ and $\tilde{\sigma}_t$. Notably, Eq.~\eqref{eq: correlation QQ} and Eq.~\eqref{eq: correlation Q0} remain distinct even at equilibrium $\rho_a = \rho_b$, unlike the corresponding single-time statistics, which coincide in that limit.
\end{subequations}

The time dependence in Eq.~\eqref{eq:correlations}, via the scaling parameters defined in Eqs.~\eqref{eq: tilda_sigma} and~\eqref{eq: sigma}, neatly captures all dynamical regimes of the active system. Moreover, the explicit dependence of Eq.~\eqref{eq:correlations} on $t_1$ resembles ageing in disordered media~\cite{2011_Berthier_Theoretical}.
 
In the rest of the \emph{Letter}, we derive these many-body results from the underlying microscopic (single-particle) dynamics.

\section{Derivation}\label{sec:single-time}
We follow an analysis previously developed in~\cite{2015_sadhu_large} for passive particles. Let $R_t$ denote the total number of AOUPs that started on the left-hand side of the origin ($x_0 \leq 0$) and are on the right-hand side of the origin ($x_t > 0$) at time $t$. Similarly, let $L_t$ denote the total number of AOUPs that started on the right-hand side of the origin ($x_0 >0$) and are on the left-hand side ($x_t\leq 0$) at time $t$. The time-integrated current up to time $t$ is then given by $Q_t = R_t - L_t$~\cite{2015_sadhu_large, 2020_Banerjee_Current}.

Since the particles are independent, the generating function, averaged over the trajectories, factorises as
\begin{subequations}\begin{equation}\label{eq: evolution average}
    \langle\mathrm{e}^{\lambda Q_t}\rangle = \langle\mathrm{e}^{\lambda R_t}\rangle \langle e^{-\lambda L_t}\rangle = \prod_{i}F_t(\lambda,y_i,u_i)
\end{equation}
where $F_t(\lambda, y, u)$ represents the contribution of a single AOUP, starting at position $y\equiv x_0$ with initial velocity $u \equiv  v_0$, to the total current. It is given by
\begin{equation}
    F_t(\lambda,y,u)=
    \begin{cases}
        1 + (e^\lambda -1) \int_{0}^{\infty} \mathrm{d}  x \, g_{t}(x |y, u) & \text{ $y\leq0$}\\
        1 + (e^{-\lambda} -1) \int_{-\infty}^{0} \mathrm{d}  x \, g_{t}(x |y, u) & \text{ $y>0$}
    \end{cases}
\end{equation}
where $g_t(x \vert y, u)$ is the single-particle propagator to find the particle at position $x$ at time $t$, regardless of its final velocity.\end{subequations}

In the annealed-position, annealed-velocity ensemble:
\begin{equation} \label{eq: annealed position average}
\dashoverline{\overline{\langle \mathrm{e}^{\lambda Q_t}\rangle }} =
    \!\! \prod_{y} \! \left\{ 1 - \rho_y \, \mathrm{d}y+\rho_y \, \mathrm{d}y F_t(\lambda, y) \right\}
\end{equation}
where $1-\rho_{y} \, \mathrm{d}y$ is the probability of finding an empty infinitesimal interval $\mathrm{d}y$, while $\rho_y \,\mathrm{d}y$ is the probability that the interval contains exactly one particle; an empty interval contributes a factor of unity to the product in Eq.~\eqref{eq: evolution average}, while $F_t(\lambda, y)= \int \dd u \,p(u)\, F_t(\lambda, y, u)$ is the velocity-averaged contribution when the interval is occupied. This gives the CGF defined in Eq.~\eqref{eq:ensemble_defs}:
\begin{equation}\label{eq: annealed annealed average}
    \mu_{\A, \A}(\lambda)  = \int_{-\infty}^{\infty} \mathrm{d}y \,\rho_y\big( F_t(\lambda, y) - 1 \big)
\end{equation}
with
\begin{align}\label{eq:F g tilde}
    F_t(\lambda,y)=
    \begin{cases}
        1 + (e^\lambda -1) \int_{0}^{\infty} \mathrm{d}  x \, \tilde{g}_{t}(x \vert y) & \text{ $y\leq0$}\\
        1 + (e^{-\lambda} -1) \int_{-\infty}^{0} \mathrm{d}  x \, \tilde{g}_{t}(x \vert y) & \text{ $y>0$}
    \end{cases}
\end{align}
in terms of the velocity-averaged single-particle propagator $\tilde{g}_{t}(x\vert y) = \int \dd u \, p(u)g_t(x\vert y, u)$. 

For the annealed-position, quenched-velocity ensemble, a similar analysis gives
\begin{equation}\label{eq: A Q average}
     \dashoverline{\ln \overline{\langle \mathrm{e}^{\lambda Q_t}\rangle }} =\int_{-\infty}^{\infty} \mathrm{d}u\,p(u) \int_{-\infty}^{\infty} \mathrm{d}y \,\rho_y\,\left( F_t(\lambda,   y, u) - 1 \right) 
\end{equation}
which is identical to Eq.~\eqref{eq: annealed annealed average}, establishing $\mu_{\A, \Q}(\lambda)=\mu_{\A, \A}(\lambda)$.

For the quenched-position, annealed-velocity ensemble, we get
\begin{equation}\label{eq: Q A average}
     \mu_{\Q, \A}(\lambda)=\overline{\ln \dashoverline{\langle \mathrm{e}^{\lambda Q_t}\rangle }} =\int_{-\infty}^{\infty} \mathrm{d}y \,\rho_y\,\ln F_t(\lambda,   y)
\end{equation}
while, for the quenched-position, quenched-velocity ensemble, 
\begin{equation}\label{eq: Q Q average}
     \mu_{\Q, \Q}(\lambda)\!=\!\overline{\dashoverline{\ln \langle \mathrm{e}^{\lambda Q_t}\rangle }} \!=\!\!\!\int_{-\infty}^{\infty} \!\!\!\!\!\!\mathrm{d}y \,\rho_y\!\!\!\int_{-\infty}^{\infty} \!\!\!\!\!\!\mathrm{d}u\,p(u)\ln F_t(\lambda, y, u)
\end{equation}

Importantly, this framework holds for any choice of single-particle active dynamics: the ensemble-averaging structure above depends only on the initial state, not on the specific microscopic propagator $g_t(x|y,u)$. The AOUP-specific results reported in this \emph{Letter} follow once we specify $g_t$; the initial state remains the domain-wall configuration $\rho_y=\rho_a$ for $y < 0$ and $\rho_b$ for $y>0$, with initial velocities drawn from Eq.~\eqref{eq: Maxwell Boltzmann}.
The explicit results in Eqs.~\eqref{eq: result aa}, \eqref{eq: result mu_qa}, \eqref{eq: 1t_mu_QQ}, and \eqref{eq: result mu_q, u=0} are obtained using the single-AOUP propagator (derived in the Supplemental Material~\cite{Supp_Mat})
\begin{equation} \label{eq: Single Particle Propagator}
    g_t(x\vert y, u) = \frac{1}{\sqrt{2 \pi \sigma_t^2}}\,\mathrm{e}^{-\frac{\left(x- b_t\right)^2}{2 \sigma_t^{2}}}
\end{equation}
where $b_t = y+u \,\tau \big(1-\mathrm{e}^{-\tfrac{t}{\tau}}\big)$ is the mean position at time $t$, which depends on the initial position $y$ and velocity $u$, and $\sigma^2_t$, given in Eq.~\eqref{eq: sigma}, is the corresponding variance. Averaging over the initial velocity distribution~\eqref{eq: Maxwell Boltzmann} yields the propagator
\begin{equation} \label{eq: tilde propagator}
    \tilde{g}_t(x\vert y) = \frac{1}{\sqrt{2 \pi \tilde{\sigma}_t^2}} \,\mathrm{e}^{-\frac{(x-y)^2}{2 \tilde{\sigma}_t^2}}
\end{equation}
where the variance $\tilde\sigma^2_t$, given in Eq.~\eqref{eq: tilda_sigma}, has a different time dependence from $\sigma_t^2$, as illustrated in Fig.~\ref{fig:one_time_probability_distribution}(c).  

Finally, Eq.~\eqref{eq: result mu_q, u=0} for the quenched-position, zero-velocity ensemble follows from Eq.~\eqref{eq: Q Q average} by setting $p(u) = \delta(u)$. 

\section{Two-time statistics} 

The two-time generating function factorises in the same way as in Eq.~\eqref{eq: evolution average}:
\begin{equation}\label{eq:t12 gf}
    \average{\mathrm{e}^{\lambda_1 Q_{t_1} + \lambda_2 Q_{t_2} }}  = \prod_{i} F_{\boldsymbol{t}}(\boldsymbol{\lambda}, y_i, u_i)
\end{equation}
with $\boldsymbol{t} \equiv (t_1, t_2)$, $\boldsymbol{\lambda} \equiv (\lambda_1, \lambda_2)$, and  $F_{\boldsymbol{t}}(\boldsymbol{\lambda}, y, u)$ the two-time analogue of $F_t(\lambda, y, u)$. For $y \leq 0$, it is given by 
\begin{align}\label{eq:F12}
    F_{\boldsymbol{t}}(\boldsymbol{\lambda},  y, u)&  =   -1+F_{t_1}(\lambda_1,y,u) + F_{t_2}(\lambda_2,y,u) \notag \\
    + & \big( e^{\lambda_1}  -1\big) \big( e^{\lambda_2} -1\big)  \int_{0}^{\infty}\mathrm{d}\boldsymbol{x }\, g_{\boldsymbol{t}}(\boldsymbol{x}\vert y,u)
\end{align}
where $\boldsymbol{x} \equiv (x_1, x_2)$ and $g_{\boldsymbol{t}}(\boldsymbol{x}|y,u)$ is the two-time single-particle propagator, giving the joint probability of being at $x_1$ at time $t_1$ and $x_2$ at time $t_2$, starting from $(y,u)$. For $y > 0$, the corresponding expression is obtained from Eq.~\eqref{eq:F12} by replacing $\lambda \to-\lambda$ and $\int_{0}^{\infty} \to \int_{-\infty}^{0}$, only in the second line.

By the Markovianity of the phase-space dynamics, 
\begin{equation*}  
g_{t_1, t_2}(x_1, x_2, v_1, v_2 \vert y, u)  = g_{t_{21}}(x_2, v_2 \vert x_1, v_1) \, g_{t_1}(x_1, v_1 \vert y, u)
\end{equation*}  
with $t_{21}=t_2-t_1$. Integrating this over the velocities $v_1$ and $v_2$ gives $g_{\boldsymbol{t}}(\boldsymbol{x}|y,u)$ in Eq.~\eqref{eq:F12} as a bivariate Gaussian,
\begin{subequations}\begin{equation}\label{eq:bi_variate_gauss}
    g_{\boldsymbol{t}}(\boldsymbol{x}\vert y,u) = \frac{1}{\sqrt{(2\pi)^2\det \boldsymbol{\Sigma}}} \mathrm{e}^{-\frac{1}{2}(\boldsymbol{x}-\boldsymbol{b_{\boldsymbol{t}}})^{\mathsf T}\boldsymbol{\Sigma}^{-1}(\boldsymbol{x}-\boldsymbol{b_{\boldsymbol{t}}})}
\end{equation}
with mean $\boldsymbol{b}_{\boldsymbol{t}} \equiv (b_{t_1}, b_{t_2})$ and $2\times2$ covariance matrix $\boldsymbol{\Sigma}_{\boldsymbol{t}}$ with diagonal elements $\Sigma_{1,1} = \sigma_{t_1}^2$, $\Sigma_{2,2} = \sigma_{t_2}^2$, and
\begin{equation}\label{eq:R12} 
\Sigma_{1,2} = \Sigma_{2,1} = \sigma_{t_1}^2+ a_{t_1}^2  \left( 1- \mathrm{e}^{-\frac{t_{21}}{\tau}}\right)
\end{equation}\end{subequations}

The two-time CGFs in different ensembles generalise their one-time counterparts in Eq.~\eqref{eq:ensemble_defs} and are expressed in terms of Eq.~\eqref{eq:F12}. For example, the quenched-position, quenched-velocity CGF becomes
\begin{equation}\label{eq: Q Q average two-time}
     \mu_{\Q, \Q}(\boldsymbol{\lambda})=\!\!\int_{-\infty}^{\infty} \!\!\!\mathrm{d}y \,\rho_y\int_{-\infty}^{\infty} \!\!\!\mathrm{d}u\,p(u)\ln F_{\boldsymbol{t}}(\boldsymbol{\lambda},  y, u)
\end{equation}
generalising Eq.~\eqref{eq: Q Q average}. Similarly, the anneald-position, annealed velocity CGF in Eq.~\eqref{eq: annealed annealed average} generalises to
\begin{equation}\label{eq: annealed annealed average two-time}
    \mu_{\A, \A}(\boldsymbol{\lambda})  = \int_{-\infty}^{\infty} \mathrm{d}y \,\rho_y\big( F_{\boldsymbol{t}}(\boldsymbol{\lambda},  y) - 1 \big)
\end{equation}
where $F_{\boldsymbol{t}}(\boldsymbol{\lambda},  y) = \int \dd u \,p(u)\, F_{\boldsymbol{t}}(\boldsymbol{\lambda},  y, u)$ is the velocity-averaged two-times contribution, analoguous to Eq.~\eqref{eq:F g tilde}. It is expressed in terms of the initial-velocity-averaged propagator $\tilde{g}_{\boldsymbol{t}}(\boldsymbol{x}\vert y)$, a bivariate Gaussian analogous to Eq.~\eqref{eq:bi_variate_gauss}, with $\boldsymbol{b}_{\boldsymbol{t}} = (y, y)$ and covariance matrix $\tilde\Sigma_{1, 1} = \tilde{\sigma}_{t_1}^2$, $\tilde\Sigma_{2, 2} = \tilde{\sigma}_{t_2}^2$, and off-diagonal elements
\begin{equation}\label{eq:tidleC12} 
    \tilde{\Sigma}_{1,2} = \tilde{\Sigma}_{2,1} = \frac{1}{2}(\tilde\sigma_{t_1}^2+ \tilde\sigma_{t_2}^2 - \tilde\sigma_{t_{21}}^2)
\end{equation}
Explicitly computing the integrals in Eq.~\eqref{eq: annealed annealed average two-time} gives the reported CGF, Eq.~\eqref{eq: 2t_mu_AA}. The structure of Eq.~\eqref{eq: 2t_mu_AA} follows directly from Eq.~\eqref{eq:F12}: the coupling term $\nu(\lambda_1,\lambda_2)$ originates from the $ g_{\boldsymbol{t}}(\boldsymbol{x}\vert y,u)$ term therein. 
A similar analysis applies to the remaining ensembles, with their results given in the Supplemental Material~\cite{Supp_Mat}.

\section{Multi-time statistics} The formalism straightforwardly extends to multi-time statistics. See~\cite{2015_sadhu_large,2026_Chowdhury} for earlier applications in passive dynamics. The multi-time generating functional
\begin{equation}\label{eq:t123 gf}
    \langle \mathrm{e}^{\int_0^T\mathrm{d}t \lambda_t Q_{t}}\rangle  = \prod_{i} F(\lambda_t, y_i, u_i)
\end{equation}
with fugacity function $\lambda_t$ and arbitrary time parameter $T$, factorises in terms of the single-particle property
\begin{equation}\label{eq:Ft pw}
      F(\lambda_t,y,u)=
    \begin{cases}
        \langle \mathrm{e}^{\int_0^T\mathrm{d}t \lambda_t \Theta(x_t)}\rangle_{y,u} & \text{ $y\leq0$}\\
        \langle \mathrm{e}^{-\int_0^T\mathrm{d}t \lambda_t \Theta(-x_t)}\rangle_{y,u} & \text{ $y>0$}
    \end{cases}
\end{equation}
where $\Theta$ is the Heaviside step function, and the average $\langle\cdot\rangle_{y,u}$ is over a single AOUP position $x_t$, starting at $y$ with velocity $u$. Averaging over the initial ensembles proceeds as before. For example, in the annealed-position, annealed-velocity ensemble, the CGF takes the same form as Eq.~\eqref{eq: annealed annealed average two-time}, with $F_{\boldsymbol{t}}(\boldsymbol{\lambda},y) \equiv \int du \, p(u)\, F_{\boldsymbol{t}}(\boldsymbol{\lambda},y, u) $ given by the series solution
\begin{equation}
   F_{\boldsymbol{t}}(\boldsymbol{\lambda},y)=1+\!\!\sum_{n\ge 1}\!\int_{0}^T\!\!\!\!\!\cdots \!\!\!\int_{t_{n-1}}^T\!\!\!\!\!\!\!\!\mathrm{d}\boldsymbol{t}\lambda(t_1)\cdots\!\lambda(t_n)\!\! \int_{0}^{\infty}\!\!\!\!\!\mathrm{d}\boldsymbol{x }\tilde{g}_{\boldsymbol{t}}(\boldsymbol{x}\vert y)
\end{equation}
for $y\le 0$ (with the corresponding expression for $y>0$ obtained as in Eq.~\eqref{eq:F12}), where $\boldsymbol{t}\equiv(t_1, ..., t_n)$, $\boldsymbol{x}\equiv (x_1,...,x_n)$, and  $\tilde{g}_{\boldsymbol{t}}(\boldsymbol{x}\vert y)$ is a multi-time generalisation of the velocity-averaged propagator in the two-time case.

\section{Conclusion}\label{Sec: Summery}

This \emph{Letter} shows how the much-studied current statistics in passive transport extend in the presence of activity. The additional scale due to persistence in single-particle dynamics introduces an intermediate regime between asymptotic diffusive regimes at short and long times (see Fig.~\ref{fig:one_time_probability_distribution}(c)). Our analysis of AOUPs neatly demonstrates how this single-particle character manifests non-trivially in the collective transport. Notably, all three dynamical regimes of current fluctuations appear through a single time-dependent scaling parameter $r_t$, which depends on the details of initial fluctuations. This sensitivity, even at long times, reflects an unusually slow relaxation of collective modes, as noted earlier in the passive context~\cite{2009_Derrida_Current,Leibovich2013,2026_saha_universal}. This long-memory effect is reflected in the slow decay of the two-time correlations that follow from our multi-time statistics. 

A point worth re-stressing is that a quenched average is different from fixing the inital state to a typical configuration.
In passive systems, the two have often been used synonymously, without it being appreciated that they can, in fact, differ. This difference is demonstrated in our results Eqs.~\eqref{eq: 1t_mu_QQ}, and \eqref{eq: result mu_q, u=0}, and is deeply rooted in the scaling of velocity fluctuations compared to that of number-density fluctuations.

The gas of independent AOUPs studied here is a simple example of active matter where many of the collective transport properties can be rigorously characterized. Despite its simplicity, it captures collective effects of activity that remain only partially understood in related, more complex systems~\cite{2020_Banerjee_Current,2023_jose_current,2024_Jose_Effect}. A major challenge is to understand how many of our observations are generic to gases of interacting active particles, where rigorous results for full-counting statistics remain extremely challenging to obtain with existing tools. Experimental realisation of some of the effects reported here would therefore be highly valuable. On the theoretical side, approaches based on fluctuating hydrodynamics~\cite{2023_agranov_macroscopic,2024_Jose_Effect} are promising, and our results would provide a benchmark for these future developments.

\acknowledgments
We thank Kapil Sharma for fruitful discussions. SJ and TS acknowledge financial support from
the Department of Atomic Energy, Government of India, under Project Identification Number RTI-4012. Work of TS was supported in part by grant NSF PHY-2309135 to the Kavli Institute for Theoretical Physics (KITP) and from the CNRS International Research Project ``Classical and quantum dynamics in out of equilibrium systems.'' We
thank the Department of Theoretical Physics, TIFR,
Mumbai, for providing computational facilities, and Ajay Salve and Kapil Ghadiali for computational support.

\bibliographystyle{eplbib}
\bibliography{refs}

\end{document}


\title{Supplement to `Current fluctuations in a gas of active Ornstein--Uhlenbeck particles'}

\author{Sandeep Jangid}
\affiliation{Department of Theoretical Physics, Tata Institute of Fundamental Research, Homi Bhabha Road, Mumbai 400005, India.}

\author{ Aman Kumbhakar}
\affiliation{Department of Theoretical Physics, Tata Institute of Fundamental Research, Homi Bhabha Road, Mumbai 400005, India.}

\author{Juliane U. Klamser}
\affiliation{Laboratoire Charles Coulomb (L2C), Universit\'e de Montpellier and CNRS (UMR 5221), 34095 Montpellier, France.}

\author{Tridib Sadhu}
\affiliation{Department of Theoretical Physics, Tata Institute of Fundamental Research, Homi Bhabha Road, Mumbai 400005, India.}

\begin{abstract}
In this Supplemental Material, we first describe the importance sampling algorithms used to obtain the rare-event simulation results shown in the \emph{Letter}. We then give the expression for the single-particle propagator, followed by the explicit two-time cumulant generating functions for the remaining initial ensembles, whose derivations follow the analysis outlined in the main text. Finally, we compare the resulting two-time distributions with our simulation results.
\end{abstract}

\maketitle

\tableofcontents

\section{Importance Sampling: One-Time }\label{Numerical_methods}

We outline the importance sampling algorithm~\cite{2023_Bello_Current,2020_Banerjee_Current,2004_hartmann_new} used to compute the full probability distribution of the integrated current $Q_t$ up to time $t$ in a one-dimensional system of non-interacting particles.

\emph{Direct sampling:} Consider a system of $N$ non-interacting particles initially confined to the interval $[-L,L]$, with particle number densities $\rho_a$ and $\rho_b$ on the left and right sides of the origin, respectively. Each particle $j$ contributes to the integrated current across the origin in one of three possible ways: (i) a particle initially on the left side ($x<0$) crosses to the right with probability $p_j^+$, contributing $+1$ to the current; (ii) a particle initially on the right side crosses to the left with probability $p_j^-$, contributing $-1$; or (iii) the particle remains on the same side of the origin with probability $p_j^0=1-p_j^+-p_j^-$, contributing zero. The corresponding single-particle current contribution $I_j$ is given by
\begin{equation}
    I_j =
    \begin{cases}
        +1, & \text{with probability} \quad p_j^+ \\
        -1, &\text{with probability} \quad p_j^- \\
        0, & \text{otherwise}
    \end{cases}
\end{equation}
The integrated current is obtained as $Q_t=\sum_{j=1}^N I_j$. Repeating this procedure over typically $10^6$--$10^7$ realizations gives a numerical estimate of the probability distribution $P(Q_t)$.

\emph{Importance sampling:} Direct sampling efficiently probes the typical region of the distribution but becomes increasingly inefficient in the tails, where the corresponding events occur with very small probabilities. To access these rare fluctuations, we employ importance sampling~\cite{2020_Banerjee_Current,2023_Bello_Current,2004_hartmann_new}, in which the single-particle probabilities are exponentially biased by a parameter $\beta$. The biased probabilities are given by
\begin{subequations}\label{eq: biased_prob}
\begin{equation}
p_j^{+,\beta} = \frac{p_j^+ e^\beta}{z_j} \qquad
p_j^{-,\beta} = \frac{p_j^- e^{-\beta}}{z_j} \qquad
p_j^{0,\beta} = \frac{1-p_j^+-p_j^-}{z_j}
\end{equation}
where $z_j = p_j^+ e^\beta + p_j^- e^{-\beta} + 1-p_j^+-p_j^-$.
\end{subequations}
The current $Q_t$ is then sampled using the biased probabilities $(p_j^{+,\beta},p_j^{-,\beta},p_j^{0,\beta})$. The biased distribution $P_\beta(Q_t)$ is related to the original distribution $P(Q_t)$ through~\cite{2023_Bello_Current}
\begin{equation}\label{eq: bias_relation}
P(Q_t)=Z e^{-\beta Q_t}P_\beta(Q_t)
\end{equation}
where $Z=\prod_{j=1}^N z_j$ is the normalization factor.

Different values of $\beta$ allow different regions of the current distribution to be sampled efficiently. Positive values of $\beta$ favor fluctuations towards the right tail, whereas negative values favor fluctuations towards the left tail. Increasing $|\beta|$ progressively shifts the sampled region further into the corresponding tail. For each value of $\beta$, the biased distribution $P_\beta(Q_t)$ is reweighted using Eq.~\eqref{eq: bias_relation} to recover the corresponding portion of the original distribution $P(Q_t)$. This procedure allows us to access probabilities over many orders of magnitude that would be practically inaccessible through direct sampling.

Thus, the implementation of both direct and importance sampling reduces to determining the single-particle probabilities $p_j^+$ and $p_j^-$, which depend on the choice of initial ensemble. In the following subsections, we provide explicit expressions for these probabilities for each ensemble.

\textbf{Annealed-position, Annealed-velocity ensemble:}
For the annealed-position, annealed-velocity ensemble, both the initial positions and initial velocities are independently sampled from their prescribed distributions. In particular, each particle is first assigned to the left or right side of the origin with probabilities $\frac{\rho_a}{\rho_a+\rho_b}$ and $\frac{\rho_b}{\rho_a+\rho_b}$, respectively, and its position is then drawn uniformly from the corresponding interval $[-L,0]$ or $[0,L]$. Independently, the initial velocity of each particle is drawn from the Maxwell–Boltzmann distribution $p(u)$ given by (2) of the \emph{Letter}.

The probabilities that a particle contributes positively or negatively to the time-integrated current are given by
\begin{equation}
p^+_j = \frac{\rho_a}{\rho_a + \rho_b} \int_{-\infty}^{\infty} \mathrm{d}u  p(u) \int_{-L}^{0} \frac{\mathrm{d}y}{L}  \int_{0}^{\infty} \mathrm{d}x  g_t(x|y,u) \quad
{\rm and} \quad
p^-_j = \frac{\rho_b}{\rho_a + \rho_b} \int_{-\infty}^{\infty} \mathrm{d}u  p(u) \int_{0}^{L} \frac{\mathrm{d}y}{L} \int_{-\infty}^{0} \mathrm{d}x \, g_t(x|y,u)\label{AA:positive_probability}
\end{equation}

Here, the propagator $g_t(x \vert y, u)$ is given in (21) of the \emph{Letter}. Since all particles are sampled from the same position and velocity distributions, the crossing probabilities $p^+ \equiv p_j^+$ and $p^- \equiv p_j^-$ are identical for all particles. The importance sampling method is then carried out by choosing the bias parameter $\beta$ following~\eqref{eq: biased_prob} and~\eqref{eq: bias_relation}.

\textbf{Quenched-position, Annealed-velocity ensemble:}
In the quenched-position-annealed-velocity ensemble, the particles are placed deterministically and equidistantly on either side of the origin, with spacings $\rho_a^{-1}$ and $\rho_b^{-1}$ on the left and right, respectively whereas the initial velocities are sampled independently from the MB distribution $p(u)$.

For a particle initially located at $y_j<0$ and $y_j>0$, respectively, the crossing probabilities
\begin{equation}
p_{y_j<0}^+(t)=
\int_{-\infty}^{\infty}\mathrm{d}up(u)
\int_{0}^{\infty}\mathrm{d}xg_t(x|y_j,u)
\qquad 
\mathrm{and} \quad
p_{y_j>0}^-(t)=
\int_{-\infty}^{\infty}\mathrm{d}up(u)
\int_{-\infty}^{0}\mathrm{d}xg_t(x|y_j,u)
\end{equation}

In contrast to the annealed-position, annealed-velocity ensemble, the crossing probabilities depend on the initial position $y_j$ and are therefore evaluated separately for each particle. 

\textbf{Annealed-position, Quenched-velocity ensemble:}
In the annealed-position, quenched-velocity ensemble, the initial positions are sampled as in the annealed-position-annealed-velocity ensemble. However, the initial velocities are quenched, hence they are fixed to a typical configuration corresponding to the MB distribution $p(u)$.

To construct this typical velocity configuration, we use inverse transform sampling. We choose $N$ uniformly spaced points $k_j=\frac{j}{N+1}$, with $j=1,2,\ldots,N$, in a unit interval $(0,1)$ and define the CDF of the MB distribution as
\begin{equation} \label{eq: u_array_1}
    \mathrm{CDF}(u)=\int_{-\infty}^{u}p(v)\,\mathrm{d}v
\end{equation}
The velocity $u_j$ of the $j$-th particle is obtained from
\begin{equation} \label{eq: u_array_2}
    \mathrm{CDF}(u_j)=k_j
\end{equation}
The resulting deterministic set $\{u_1,u_2,\ldots,u_N\}$ provides a typical velocity configuration representative of the MB distribution.

The probabilities of contributing positively and negatively to the integrated current are
\begin{align}
p_{u_j}^+(t) = \frac{\rho_a}{\rho_a + \rho_b} \int_{-L}^{0} \mathrm{d}y \, \frac{1}{L} \int_{0}^{\infty} \mathrm{d}x \, g_t(x|y,u_j)\qquad
{\rm and }\qquad
p_{u_j}^-(t) = \frac{\rho_b}{\rho_a + \rho_b} \int_{0}^{L} \mathrm{d}y \, \frac{1}{L} \int_{-\infty}^{0} \mathrm{d}x \, g_t(x|y,u_j)
\end{align}
Similar to the quenched-position-annealed-velocity ensemble, the crossing probabilities depend on initial velocity $u_j$. Therefore, probabilities $p_{u_j}^+$ and $p_{u_j}^-$ are evaluated separately for each particle and subsequently used in the importance sampling procedure.

\textbf{Quenched-position, Quenched-velocity ensemble:} 
For the quenched-position, quenched-velocity ensemble, both the initial positions and velocities are fixed following the procedures described above. The velocities obtained from Eqs.~\eqref{eq: u_array_1} and~\eqref{eq: u_array_2} are  assigned randomly to the particles with fixed positions. 
The crossing probabilities $p_{y_j, u_j}^+$ and $p_{y_j, u_j}^-$ are given by

\begin{equation}
    p_{y_j, u_j}^+ = \int_{0}^{\infty} \mathrm{d}x \, g_t(x | y_j, u_j)\qquad
{\rm and} \qquad
    p_{y_j, u_j}^- = \int_{-\infty}^{0} \mathrm{d}x \, g_t(x | y_j, u_j)
\end{equation}
Thus, the probabilities are evaluated separately for each fixed pair $(y_j,u_j)$.

\textbf{Quenched-position, zero-velocity ensemble:}
For the quenched-positions, zero-velocity ensemble, the particles are placed at the fixed positions as discussed in the quenched-position ensembles, and the initial velocity is set to $u_j=0$ for every particle. The corresponding crossing probabilities are 
\begin{equation}
p_{y_j}^+(t) = 
\int_{0}^{\infty} \mathrm{d}x \, g_t(x|y_j) \quad
{\rm and } \quad
p_{y_j}^-(t) = \int_{-\infty}^{0} \mathrm{d}x \, g_t(x|y_j)
\end{equation}
In this case, the probabilities depend only on the fixed initial position $y_j$ of each particle.

The propagators entering the expressions above are given in the \emph{Letter}, and the corresponding integrals are evaluated numerically to the desired precision.

\section{Importance sampling: Two-time } 
\label{sec:two_time_sampling}

We extend the importance sampling algorithm to the two-time probability distribution of the currents $Q_{t_1}$ and $Q_{t_2}$, to sample rare events in the joint distribution of the currents measured at times $t_1$ and $t_2$, respectively. 

While in the single-time case each particle contributes to the current through one of three possible scenarios ($+$, $-$, or $0$), the two-time case involves seven possible scenarios $\{(+,+), (0, +), (+,0), (-,0), (0,-), (-,-),(0,0) \}$ which specify each particle’s contribution at the two times. These joint transition probabilities of the particle are denoted by $p_j^{m,n}$, where $m, n \in \{+, -, 0\}$. For example, $p_j^{+,0}$ represents the probability that a particle initially on the left of the origin contributes to positive current at $t_1$ (found on the right side at $t_1$) and zero current at $t_2$ (found on the left side at $t_2$).

Similar to the single-time case, the indicator function for each $j$-th particle becomes two-dimensional
\begin{equation}\label{eq:two_time_indicator}
   \mathbf{I}_j = 
   \begin{cases} 
   (1,1) & \text{with probability } \quad p_j^{+,+} \\
   (-1,-1) & \text{with probability } \quad  p_j^{-,-} \\
   \vdots & \vdots \\
   (0,0) & \text{otherwise}
   \end{cases}
\end{equation}
where $\mathbf{I}_j$ represents the particle's contribution at times $(t_1, t_2)$. The total joint currents are given by $(Q_{t_1},Q_{t_2}) = \sum_{j=1}^N \textbf{I}_j$.

The generalization of importance sampling to two times is implemented by introducing two exponential biasing parameters $\beta_1$ and $\beta_2$, which modify the probabilities as
\begin{equation}
p_j^{(m,n),(\beta_1,\beta_2)} = \frac{1}{z_j}p_j^{m,n}e^{m\beta_1 + n\beta_2}
\end{equation}
where, in the exponential factor, $+,-,$ and $0$ are identified with $+1,-1,$ and $0$, respectively, and the normalization factor
\begin{equation}
z_j = \sum_{m,n}p_j^{m,n}e^{m\beta_1 + n\beta_2}
\end{equation}
The parameters $\beta_1$ and $\beta_2$ control the regions of the joint $(Q_{t_1},Q_{t_2})$ distribution that are preferentially sampled. Different choices of $(\beta_1,\beta_2)$ allow us to probe different regions of the $(Q_{t_1},Q_{t_2})$ plane. For each pair $(\beta_1,\beta_2)$, the biased histogram obtained by using the biased probabilities in the indicator function \eqref{eq:two_time_indicator} can be re-weighted to the corresponding original joint distribution as

\begin{equation}
P_{t_1,t_2}(Q_{t_1},Q_{t_2}) = Ze^{-\beta_1 Q_{t_1} - \beta_2 Q_{t_2}}P_{t_1,t_2}^{\beta_1,\beta_2}(Q_{t_1},Q_{t_2})
\label{eq:biased_to_unbiased}
\end{equation}

where, $Z = \underset{j}{\prod} z_j$. These bias parameters are chosen such that the distributions obtained for different pairs overlap over suitable regions of $(Q_{t_1},Q_{t_2})$. These overlapping portions agree with each other and are combined to reconstruct the full two-time distribution, including its far tails.

For AOUPs, the transition probabilities $p_j^{m,n}$ are calculated using the joint propagator given in (25) of the \emph{Letter}, which depends on the particle's initial conditions. For all initial ensembles, we follow the discussion in Sec.~\ref{Numerical_methods} to calculate these transition probabilities $p_j^{m,n}$.

\section{The AOUP Propagator}

The phase-space propagator $g_t(x, v \vert y, u)$ of an AOUP, representing its probability to evolve from an initial state $(y, u)$ at $t=0$ to a state $(x, v)$ at time $t$, is given by~\cite{2021_bothe_doi, 2021_Martin_statistical}
\begin{subequations}\label{eq: Joint Propagator}\begin{equation}
    g_t(x, v \vert y, u) = \frac{\exp\left(-\frac{\left(x - y - (u + v)\theta_t \right)^2}{2 \beta_t ^2}\right)}{\sqrt{2\pi \beta_t ^2}}  
     \frac{\exp\left(-\frac{\left(v - u \mathrm{e}^{-t/\tau}\right)^2}{2 \omega_t^2}\right)}{\sqrt{2\pi \omega_t^2}}
\end{equation}
where
\begin{align}
    \theta_t &=  \tau \tanh\left( \frac{t}{2 \tau} \right) \label{eq: nu_t}\\
    \beta_t^{2} &= 2 D t \left( 1 + {\rm Pe}^2 \right) - 4 D \theta_t {\rm Pe}^2  \label{eq: gamma_t} \\   
    \omega_t^2 &= \frac{D}{\tau} {\rm{Pe}^2} \left(1 - \mathrm{e}^{-2t/\tau} \right)\label{eq: kappa_t}
\end{align}\end{subequations}
The propagator is straightforward to obtain by solving the corresponding Fokker-Planck equation.

The stationary probability distribution of the velocity given in (1c) of the \emph{Letter} is obtained after integrating~\eqref{eq: Joint Propagator} over position, $g_t(v |y, u) = \int \mathrm{d} x \, g_t(x, v | y, u)$ and taking the stationary limit $p(v) = \underset{t \rightarrow \infty}{\lim} g_t(v|y, u)$. Similarly, integrating~\eqref{eq: Joint Propagator} over $v$ gives the position propagator $g_t(x|y,u)$ given in (21) of the \emph{Letter}, while averaging over the stationary initial-velocity distribution $p(u)$ gives the velocity-averaged propagator $\widetilde{g}_t(x|y)$ in (22) of the \emph{Letter}.

\section{Two-time current statistics} \label{Sec: Two-Time correlations}

\begin{figure}
    \centering
    \includegraphics[width=1.0\textwidth]{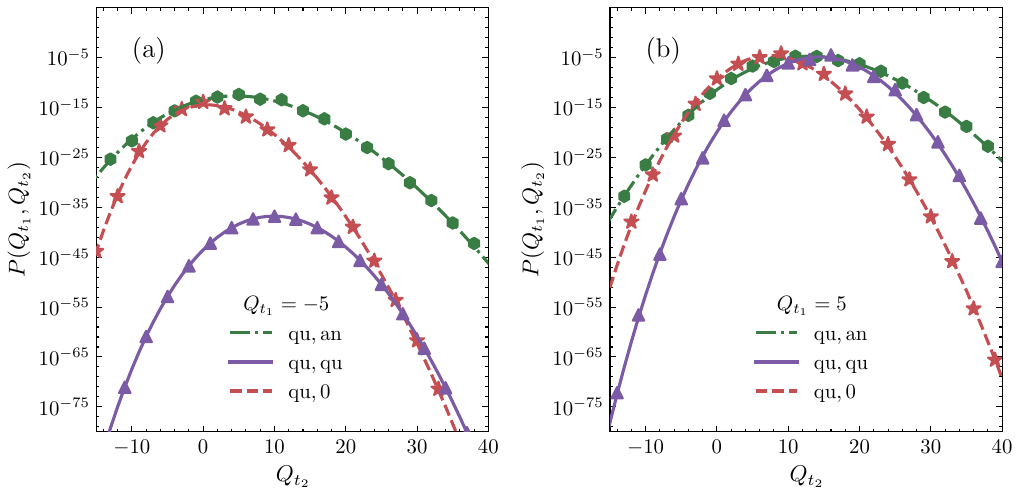}
    \caption{\emph{Domain-wall state:} Two-time current probability distribution at times $t_1 = 10$ and $t_2 = 20$ for non-uniform densities $(\rho_a, \rho_b) = (\tfrac{3}{2}, \tfrac{1}{2})$. The distributions shown are for fixed $Q_{t_1}$, with panel (a) for $Q_{t_1} = -5$ and panel (b) for $Q_{t_1} = 5$. Three initial ensembles are considered in both panels. (i) $(\Q, \A)$: the dot-dashed green line is the distribution obtained from the large-deviation asymptotic form [Eq.~(8) of the \emph{Letter}], with the LDF given by the Legendre transform of the CGF in Eq.~\eqref{eq: 2t_qa}; the green hexagons are the corresponding rare-event simulation results. (ii) $(\Q, \Q)$: the solid blue line is the theoretical result obtained from the CGF in Eq.~\eqref{eq: 2t_qq}; the blue triangles are the corresponding simulation results. (iii) $(\Q, 0)$: the dashed red line is the theoretical result obtained from the CGF in Eq.~\eqref{eq: 2t_qa}, with $\sigma_t$ and $\Sigma_t$ replacing $\tilde\sigma_t$ and $\tilde\Sigma_t$; the red stars are the corresponding simulation results.
}
    \label{fig:2t_QA}
\end{figure}

The two-time CGF of current for the annealed-position, annealed-velocity ensemble is given in Eq.~(12) of the \emph{Letter} and the corresponding numerical confirmation is shown in Fig.~3 of the \emph{Letter}. We list below the expression of the two-time CGF for the remaining ensembles and their numerical confirmation in Fig.~\ref{fig:2t_QA} and Fig.~\ref{fig:2t_same_density}.

\subsection{Annealed-position, Annealed-velocity}
The time-dependent part of the correlation term $\nu \left(\lambda_1,\lambda_2 \right)$ given in (12b) of the \emph{Letter}, requires computation of the following nested integrals
\begin{equation}  
     \int_{-\infty}^{0} \mathrm{d} y\int_{0}^{\infty}\mathrm{d}\boldsymbol{x }\, \tilde{g}_{\boldsymbol{t}}(\boldsymbol{x}\vert y) = 
    \frac{1}{2 \pi \sqrt{(1-\tilde{R}^2)\tilde{\sigma}_{t_1}^2 \tilde{\sigma}_{t_2}^2}}  
    \int_{-\infty}^{0} \!\!\mathrm{d} y \int_{0}^{\infty} \!\!\mathrm{d}x_{1} \int_{0}^{\infty} \!\!\mathrm{d}x_{2} \,  
    \mathrm{e}^{-\frac{1}{ \left(1-\tilde{R}^2\right)} \left(  
    \frac{(x_1-y)^2}{2\tilde{\sigma}_{t_1}^2} + \frac{(x_2-y)^2}{2\tilde{\sigma}_{t_2}^2} - \frac{ \tilde{R} (x_1-y)(x_2-y)}{\tilde{\sigma}_{t_1} \tilde{\sigma}_{t_2}}  
    \right)}
\end{equation}
where $\tilde{R} \equiv \frac{\tilde\Sigma_{1,2}}{\tilde{\sigma}_{t_1}\tilde{\sigma}_{t_2}}$. The $x_2$-integral is straightforward to compute, which leads to
\begin{equation}
     \int_{-\infty}^{0} \mathrm{d} y\int_{0}^{\infty}\mathrm{d}\boldsymbol{x }\, \tilde{g}_{\boldsymbol{t}}(\boldsymbol{x}\vert y) = \int_{0}^{\infty} \mathrm{d}x_{1} \int_{-\infty}^{0}  \mathrm{d}  y \frac{e^{-\frac{(x_1 -y)^2}{2 \tilde{\sigma}_{t_1}^2}} }{2 \sqrt{2 \pi  \tilde{\sigma}_{t_1}^2}}\left(1+ \text{erf}\left(\frac{\tilde{R} \tilde{\sigma}_{t_2} x_1+y (\tilde{\sigma}_{t_1}-\tilde{R} \tilde{\sigma}_{t_2})}{\sqrt{2\left(1-\tilde{R}^2\right) \tilde{\sigma}_{t_1}^2 \tilde{\sigma}_{t_2}^2} }\right)\right)
\end{equation}
Further simplification comes after a coordinate transformation $w =  x_1 - y$,
\begin{equation}  
    \int_{-\infty}^{0} \mathrm{d} y\int_{0}^{\infty}\mathrm{d}\boldsymbol{x }\, \tilde{g}_{\boldsymbol{t}}(\boldsymbol{x}\vert y) = \int_{0}^{\infty} \mathrm{d}x_{1} \int_{x_1}^{\infty} \mathrm{d} w \,  
    \frac{e^{-\frac{w^2}{2 \tilde{\sigma}_{t_1}^2}}}{2 \sqrt{2 \pi \tilde{\sigma}_{t_1}^2} }  
    \operatorname{erfc}\left(  
    \frac{w (\tilde{\sigma}_{t_1} - \tilde{R} \tilde{\sigma}_{t_2}) - \tilde{\sigma}_{t_1} x_1}{\sqrt{2(1-\tilde{R}^2) \tilde{\sigma}_{t_1}^2 \tilde{\sigma}_{t_2}^2}}  
    \right)
\end{equation}
and by change of order of the integrations,  $\int_{0}^{\infty} \mathrm{d} x_1 \int_{x_1}^{\infty} \mathrm{d} w \rightarrow \int_{0}^{\infty} \mathrm{d} w \int_{0}^{w} \mathrm{d} x_1$, the two integrals reduce into
\begin{equation}
    \int_{-\infty}^{0} \mathrm{d} y\int_{0}^{\infty}\mathrm{d}\boldsymbol{x }\, \tilde{g}_{\boldsymbol{t}}(\boldsymbol{x}\vert y) = \frac{1}{2 \sqrt{2 \pi}} \left( \tilde{\sigma}_{t_1} + \tilde{\sigma}_{t_2} - \sqrt{\tilde{\sigma}_{t_1}^2 - 2 \tilde{R} \tilde{\sigma}_{t_1}\tilde{\sigma}_{t_2} + \tilde{\sigma}_{t_2}^2}\right) 
\end{equation}
Using the expression of $\tilde{\Sigma}_{1,2}$ given in (28) of the \emph{Letter} into (27) of the \emph{Letter}, we recover the result (12) of the \emph{Letter}. Expanding the CGF in powers of $\lambda_1, \lambda_2$, we obtain the exact expression of the two-time correlation function given in (13a) of the \emph{Letter}.

\subsection{ Quenched-position, Quenched-velocity}
The CGF for the quenched-position, quenched-velocity ensemble is given by
\begin{subequations}\label{eq: 2t_qq}\begin{align}
     \mu_{\mathrm{qu}, \mathrm{qu}} =    \int_{0}^{\infty} \mathrm{d} y &\int_{-\infty}^{\infty} \frac{\mathrm{d} z }{\sqrt{ \pi }} \mathrm{e}^{- z^2}\sqrt{2} \Bigg[ \rho_a \ln \Big(1 +  \frac{\mathrm{e}^{\lambda_1}-1}{2}\operatorname{erfc}\left(h_1
    \right) +  \frac{\mathrm{e}^{\lambda_2}-1}{2}\operatorname{erfc}\left(h_2
    \right)+ \frac{(\mathrm{e}^{\lambda_1}-1)(\mathrm{e}^{\lambda_2}-1)}{4} I_{t_1, t_2}(h_1, h_2)\Big)\notag\\
    & + \rho_b \ln \Big(1 +  \frac{\mathrm{e}^{-\lambda_1}-1}{2}\operatorname{erfc}\left(h_1
    \right) +  \frac{\mathrm{e}^{-\lambda_2}-1}{2}\operatorname{erfc}\left(h_2
    \right)+ \frac{(\mathrm{e}^{-\lambda_1}-1)(\mathrm{e}^{-\lambda_2}-1)}{4} I_{t_1, t_2}(h_1, h_2)\Big)\Bigg]
\end{align}
where $h_i =  \frac{y  + a_{t_i} z }{\sigma_{t_i}}$, and
\begin{equation}
        I_{t_1, t_2}(h_1, h_2) = \operatorname{erfc}\left(h_1\right)
\operatorname{erfc}\left(h_2\right) + \frac{2}{ \pi} \int_{0}^{\tfrac{\Sigma_{12}}{\sigma_{t_1}\sigma_{t_2}}} \mathrm{d} s \frac{\exp \left( -\frac{h_1^2 - 2 s h_1 h_2 + h_2^2}{(1-s^2)}\right)}{\sqrt{1-s^2}}
\end{equation}\end{subequations}
Expanding the CGF in powers of $\lambda_1, \lambda_2$, we obtain the exact expression of the two-time correlation function given in (13c) of the \emph{Letter}.

\begin{figure}
        \centering
        \includegraphics[width=1\textwidth]{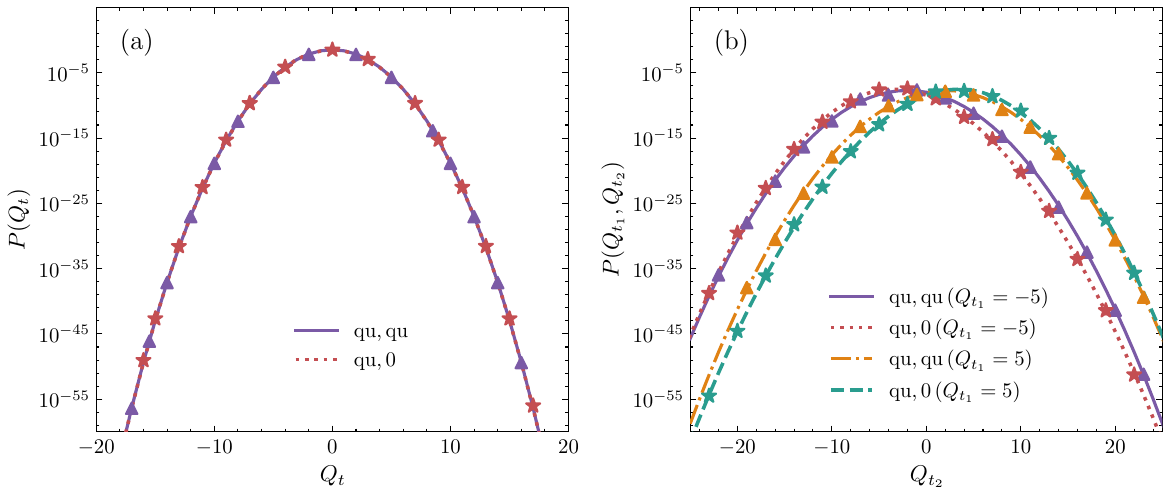}
        \caption{\emph{Equilibrium state:} (a) Comparison of the single-time current probability distribution in the $(\Q,\Q)$ and $(\Q,0)$ ensembles. The lines indicate the theoretical results obtained from the corresponding single-time CGFs, given in Eqs.~(6) and~(9) of the \emph{Letter}. The data points denote the corresponding numerical simulation results, with triangles for the $(\Q,\Q)$ ensemble and stars for the $(\Q,0)$ ensemble. (b) Two-time current probability distribution at times $t_1 = 10$ and $t_2 = 20$, shown for the fixed values of $Q_{t_1}$ indicated in the legend. For the $(\Q,\Q)$ ensemble, the theoretical predictions (lines), obtained using the Legendre transform of the CGF in Eq.~\eqref{eq: 2t_qq}, match the corresponding simulation results (triangles). For the $(\Q,0)$ ensemble, the theory (lines) likewise agrees with the simulations (stars). Both panels are for a uniform initial state with average densities $(\rho_a, \rho_b) = (1,1)$.
}
        \label{fig:2t_same_density}
\end{figure}

\subsection{Quenched-position, Annealed-velocity}
The CGF for the quenched-position, annealed-velocity ensemble is
\begin{subequations}\label{eq: 2t_qa}\begin{align}
    \mu_{\mathrm{qu}, \mathrm{an}} & = \rho_a\int_{0}^{\infty} \mathrm{d} y \ln \Big( 1 +  \frac{\mathrm{e}^{\lambda_1}-1}{2}\operatorname{erfc}\left(h_1
    \right) +  \frac{\mathrm{e}^{\lambda_2}-1}{2}\operatorname{erfc}\left(h_2
    \right)+ \frac{(\mathrm{e}^{\lambda_1}-1)(\mathrm{e}^{\lambda_2}-1)}{4} f_{t_1, t_2}(h_1, h_2)\Big) \notag\\
    & +\rho_b\int_{0}^{\infty} \mathrm{d} y \ln \Big( 1 +  \frac{\mathrm{e}^{-\lambda_1}-1}{2}\operatorname{erfc}\left(h_1
    \right) +  \frac{\mathrm{e}^{-\lambda_2}-1}{2}\operatorname{erfc}\left(h_2
    \right)+ \frac{(\mathrm{e}^{-\lambda_1}-1)(\mathrm{e}^{-\lambda_2}-1)}{4} f_{t_1, t_2}(h_1, h_2)\Big)
\end{align}
where $h_i = \frac{y}{\sqrt{2}\tilde{\sigma}_{t_i}}$
with 
\begin{align}
    f_{t_1, t_2}(h_1, h_2) = \int_{-\infty}^{0} \mathrm{d}x_{1} \frac{2 \, \mathrm{e}^{-(x_1 -h_1)^2 }}{ \sqrt{\pi  }}\left(1- \text{erf}\left(\frac{\tilde{R} \,x_1 + (h_2-h_1\tilde{R} )}{\sqrt{\left(1-\tilde{R}^2\right) } }\right)\right) \quad \text{and  } \tilde{R} = \frac{\tilde\Sigma_{12} }{\tilde\sigma_{t_1} \tilde\sigma_{t_2}} 
\end{align}
and the correlation (13b) in the \emph{Letter} can be obtained by expanding to second order in $\lambda_1$ and $\lambda_2$.
\end{subequations}

\subsection{Quenched-position, Zero-velocity ($u=0$)}
The zero velocity initial state, $u=0$, corresponds to distribution $p(u) = \delta(u)$. The resulting CGF for the quenched-position, zero-velocity ensemble yields the same formal expression \eqref{eq: 2t_qa} with $\sigma_t$ and $\Sigma_t$ replacing $\tilde\sigma$ and $\tilde\Sigma_t$.

Figure~\ref{fig:2t_same_density}(a) shows that, for a uniform initial state, the single-time CGFs of the $(\Q,0)$ and $(\Q,\Q)$ ensembles coincide. The two ensembles nevertheless differ in their multi-time statistics, even for a uniform initial state, as evidenced in Fig.~\ref{fig:2t_same_density}(b).

\bibliography{refs}